\documentclass{aa}  

\usepackage{graphicx}
\usepackage{hyperref}
\usepackage{txfonts}
\usepackage{color}
\usepackage{xcolor}
\usepackage{longtable}
\usepackage{txfonts}
\usepackage{rotating}
\usepackage{lscape}
\usepackage{booktabs}
\usepackage{multirow}
\usepackage{soul}

\begin{document}

   \title{The extended halo of NGC~2682 (M~67)}
   \subtitle{from Gaia DR2}
   %   \title{Twice as big but falling apart}
   %   \subtitle{the (disk-) shocking truth on M~67 revealed by Gaia DR2}

   \author{R. Carrera
          \inst{1}
          \and M. Pasquato\inst{1} 
          \and A. Vallenari\inst{1}
          \and L. Balaguer-N\'u\~nez\inst{2}
          \and T. Cantat-Gaudin\inst{2}
          \and M. Mapelli\inst{1,3,4,5}
          \and A. Bragaglia\inst{6}
          \and D. Bossini\inst{1,7} 
          \and C. Jordi\inst{2}
          \and D. Galad\'{\i}-Enr\'{\i}quez\inst{8}
          \and E. Solano\inst{9}
          }

   \institute{INAF-Osservatorio Astronomico di Padova, vicolo dell'Osservatorio 5, 35122 Padova (Italy)\\
              \email{jimenez.carrera@inaf.it}
    \and
Institut de Ci\`encies del Cosmos, Universitat de Barcelona (IEEC-UB), Mart\'i i Franqu\`es 1, E-08028 Barcelona, Spain  
         \and
Physics and Astronomy Department Galileo Galilei, University of Padova, Vicolo dell'Osservatorio 3, I-35122, Padova, Italy
\and 
INFN-Padova, Via Marzolo 8, I-35131 Padova, Italy
\and 
Institut f\"ur Astro- und Teilchenphysik, Universit\"at Innsbruck, Technikerstrasse 25/8, A-6020, Innsbruck, Austria
\and
    INAF-Osservatorio di Astrofisica e Scienza dello Spazio, via P. Gobetti 93/3, 40129 Bologna, Italy
         \and
           Instituto de Astrof\'isica e Ci$\hat{e}$ncias do Espa\c{c}o, Universidade do Porto, CAUP, Rua das Estrelas, 4150-762 Porto,Portugal
           \and{Observatorio de Calar Alto, Sierra de los Filabres, E-04550-G\'ergal (Almer\'{\i}a), Spain}
         \and
            Centro de Astrobiolog\'{\i}a (INTA-CSIC), Departamento de Astrof\'{\i}sica. P.O. Box 78, E-28691, Villanueva de la Ca\~nada, Madrid, Spain; Spanish Virtual Observatory
           }

   \date{Received; accepted }

% \abstract{}{}{}{}{} 
% 5 {} token are mandatory
 
  \abstract
  {NGC~2682 is a nearby open cluster, approximately 3.5\,Gyr old. Dynamically, most open clusters should dissolve on shorter timescales, of $\approx$ 1\,Gyr. Having survived until now, NGC~2682 was likely much more massive in the past, and is bound to have an interesting dynamical history.}
   %{\textcolor{blue}{Limited to 3000 words (four or five pages) but can have unlimited supporting material as appendices.}}
   {We investigate the spatial distribution of NGC~2682 stars to constrain its dynamical evolution, especially focusing on the marginally bound stars in the cluster outskirts.}
   {We use \textit{Gaia}~DR2 data to identify NGC~2682 members up to a distance of $\sim$150\,pc (10 degrees). Two methods ({\it Clusterix} and {\it UPMASK}) are applied to this end. We estimate distances to obtain three-dimensional stellar positions using a Bayesian approach to parallax inversion, with an appropriate prior for star clusters. We calculate the orbit of NGC~2682 using the \textit{GRAVPOT16} software.}
   {The cluster extends up to 200\arcmin~(50\,pc) which implies that its size is at least twice as previously believed. This exceeds the cluster Hill sphere based on the Galactic potential at the distance of NGC~2682.}
   {The extra-tidal stars in NGC~2682 may originate from external perturbations such as disk shocking or dynamical evaporation from two-body relaxation. The former origin is plausible given the orbit of NGC~2682, which crossed the Galactic disk $\approx 40$\,Myr ago.}


   \maketitle

\section{Introduction}

The dynamical evolution of a stellar cluster is strongly affected by internal collisional dynamics, i.e. two-body gravitational encounters between stars \citep[see][for an in-depth discussion in the case of globular star clusters]{1987degc.book.....S}, stellar-wind mass loss, and external interactions such as encounters with molecular clouds \citep[][]{1958ApJ...127...17S, 2009gcgg.book..375G} or Galactic disk and spiral arms \citep[][]{1987gady.book.....B}. These phenomena modify the cluster structure and internal distribution of the stars along with its lifetime and eventually bring the cluster to dissolution. In fact, they are so effective that dynamical models predict that a cluster with the typical mass of a bound open cluster (OC), $\lesssim{}10^4\,{}$M$_\odot$, would be dissolved in a timescale of $\lesssim{}1$\,Gyr \citep[see e.g.][]{2003MNRAS.340..227B, lamers2005analytical, 2008ASPC..388..353B}.

However, about 20\% of the known OCs have ages older than 1\,Gyr \citep{dias2002,kharchenko2013}. Although they are a minority, they are key to investigate the dynamical evolution of clusters, and the balance among self gravity, internal dynamics, and external tidal forces. These old OCs have several features that differentiate them from the bulk of the OCs population. They are systematically more massive than the younger bound clusters.  In particular, although these clusters are unambiguously related to the Galactic disk because of their age and chemical composition, they are also often found at high altitude above the Galactic plane, away from the disk's disruptive influence. Even though, strictly speaking, it would be hard to exclude that that some cluster have been acreeted from external galaxies in merging processes.

NGC~2682 (M~67) is one of the nearest old OCs \citep[3.6\,Gyr,][]{bossini2019} located at about 860\,pc from the Sun \citep{tristan2018b} to which it is quite similar in terms of age and initial chemical composition \citep[e.g.][]{2016MNRAS.463..696L}, to the point that it has long been debated whether the Sun might actually have originated from NGC~2682 \citep[see][and references therein]{2012AJ....143...73P}. Evidence of substantial dynamical evolution in this cluster has been reported by several authors, such as the existence of mass segregation  \citep[e.g.][]{fan1996,balaguer_nunez2007,bonatto2003,davenport2010,geller2015,gao2018} or an elongated halo roughly aligned with the cluster proper motion \citep{davenport2010}. In fact dynamical simulations of this cluster suggest that it initially was $10$ times more massive \citep{hurley2005}.

The astrometric information provided by the second data release of the \textit{Gaia} mission, \textit{Gaia}~DR2 \citep[][]{gaia_mission,GAIA_GDR2}
has proven a valuable tool to investigate the extra-tidal regions of several OCs in the solar neighborhood. Tidal tails have been reveled in the Hyades \citep[Melotte~25][]{meingast2019,roser2019}, Coma Berenices \citep[Melotte~111][]{furnkranz2019arXiv190207216F}, and Praesepe \citep[NGC~2632][]{roser2019presepe} open clusters. All of them with ages younger than 1\,Gyr \citep[e.g.][]{bossini2019}. However, hints of the existence of cluster member beyond the tidal radius have also been reported around the cluster Ruprecht~147 \citep{yeh2019} with an age older than 1\,Gyr \citep[e.g.][]{kharchenko2013}. 

In this paper we studied the current dynamical status of NGC~2682 using \textit{Gaia}~DR2. The observational material utilized in this paper is presented in Sect.~\ref{sect:data}. The obtained %NGC~2682 
density profile is discussed in Sect.~\ref{sect:profile}. The 3D spatial distribution is derived in Sect.~\ref{ReverendBayes} using a Bayesian approach to parallax inversion. Finally, in Sect.~\ref{physics} we draw conclusions, showing that NGC~2682 is more extended than expected and that its outer envelope of extra-tidal stars may be the result of disk shocking.

%%%%%%%%FIGURE%%%%%%%%
\begin{figure}
\centering
\includegraphics[scale=0.35]{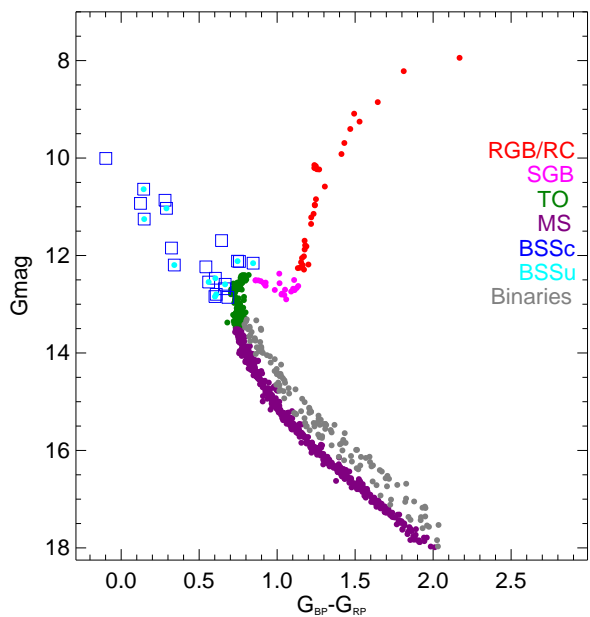}
\caption{\textit{Gaia}~DR2 colour-magnitude diagram of the stars labeled as members by both \textit{UPMASK} and \textit{Clusterix}. The different populations have been plotted in different colours: RGB/RC (red), SGB (magenta), TO (green), MS (purple), BSS from \textit{Clusterix} (BSSC, blue squares) and \textit{UPMASK} (BSSu, cyan), and binaries (grey).}
\label{fig:dcm}
\end{figure}
%%%%%%%%FIGURE%%%%%%%%

\section{The data}\label{sect:data}

\textit{Gaia}~DR2 provides $5$-parameter astrometric solution, i.e. positions ($\alpha$ and $\delta$), proper motions ($\mu_{\alpha*}$, $\mu_{\delta}$), and parallaxes ($\varpi$) plus magnitudes in three photometric bands ($G$, $G_{\rm BP}$ and $G_{\rm RP}$) for more than 1.3 billion sources, and radial velocities (RV) for more than 7 million stars \citep{GAIA_GDR2,katz18,lindegren2018}.
We study the data from \textit{Gaia}~DR2 in a radius of 10\,deg ($\sim 150$~pc) from the
 centre of NGC~2698 as determined by \citet{tristan2018b}: $\alpha_{c}$=132\fdg846 and $\delta_{c}$=11\fdg814. We follow the recommendations by \citet{arenou2018A&A...616A..17A}, \citet[][eqs. C.1 and C.2]{lindegren2018} and \textit{Gaia}~DR2 kown-issues cosmos webpage\footnote{\url{https://www.cosmos.esa.int/web/gaia/dr2-known-issues}} for an astrometrically clean subset. We also limit our analysis to those objects with low parallax uncertainties. The uncertainty in parallax depends on the signal-to-noise ratio, which is correlated with $G$ magnitude, but also on the fact that the star is a binary. Therefore we rejected those stars with error in parallaxes larger than two times the typical error at that $G$ magnitude. Finally, we constrain our analysis  to those stars with $G$ magnitude brighter than 18\,mag to avoid problems with the completeness of the sample at fainter magnitudes. We have not attempted to use the radial velocities in our analysis because \textit{Gaia} only provides radial velocities for the very bright objects \citep[$G\leq$12, see][]{katz18, soubiran2018} and other studies available in the literature \citep[e.g.][]{geller2015} cover areas smaller than that explored in this work.  

We used two different methods in order to determine the membership probabilities: \textit{UPMASK} \citep[Unsupervised Photometric Membership Assignment in Stellar Clusters;][]{upmask} and \textit{Clusterix}\footnote{Available at \url{http://clusterix.cab.inta-csic.es/clusterix/}} \citep{clusterix} codes. Although \textit{UPMASK} was originally developed to handle photometric data it can be also applied to astrometric data \citep[see the discussion in ][]{cantatgaudinupmask}. \textit{UPMASK} assigns the membership probabilities on the basis that member stars must share common properties in $\mu_{\alpha*}$, $\mu_{\delta}$, and $\varpi$ space and are more concentrated in $\alpha$-$\delta$ plane than a random distribution \citep[see][for details]{cantatgaudinupmask}. On the other hand, \textit{Clusterix} performs an empirical determination of the field and cluster frequency functions in proper motions ($\mu_{\alpha*}$, $\mu_{\delta}$) by comparing two areas: the cluster core and one far enough to represent the field population in the cluster area (see Appendix~\ref{apen:clusterix} for details). For \textit{UPMASK} we considered as cluster members those stars with an astrometric probability, $p\geq$ 0.4 \citep{tristan2018b,soubiran2018}. In the case of \textit{Clusterix} we selected as member candidates those objects with $p\geq$ 0.81 following the procedure described in Appendix~\ref{apen:clusterix}, applying after an additional constrain in $\varpi$. \textit{Clusterix} provides a slightly larger number of members than \textit{UPMASK}. Because there is no clear reason to prefer one method over the other, we selected those objects for which both methods agree in labelling them as members.  Finally, we fine-tuned the sample on the basis of the positions in the color-magnitude diagram: main-sequence (MS), turn-off (TO), sub giant branch (SGB), red giant branch (RGB), and red clump (RC). Basically, we used the objects inside a radius of 25\arcmin~to define the regions used for this purpose (Fig~\ref{fig:dcm}). After all this procedure our sample is formed by 808 objects inside a radius of 150\,pc. Following the same procedure we have selected stars redward of the main-sequence in the expected location of the binary sequence. In total there are 144 stars in this region (grey points in Fig.~\ref{fig:dcm}).

There are also several stars above the main-sequence turn-off known as blue straggler stars (BSSs, see Fig~\ref{fig:dcm}). The formation of these objects is still not well understood but the proposed scenarios include mass transfer or mergers in close binary systems \citep[e.g.][]{mccrea1964} or collisions between single stars or multiple systems that lead to a merger \citep[e.g.][]{hills1976}. They are key particles to investigate the internal dynamics because they are typically located in the central regions, probably as a consequence of dynamical friction \citep[e.g.][]{mapelli2004}. According to \citet[][and references therein]{geller2015} there are 25 BSS candidates in this cluster (see Table~\ref{tab:bss}). One of these stars, S\,1082, is not present in the \textit{Gaia}~DR2 catalogue. This is a triple system which includes a RS CVn-type eclipsing binary \citep{belloni1998}. Both \textit{Clusterix} and \textit{UPMASK} methods yield very low membership probabilities for nine of these stars. Two of them have been discarded on the basis of their radial velocities and the rest are binaries according to \citet{geller2015}. Finally we get a list with eleven BSS candidates with high membership probabilities for the two methods: two of them are new findings of this work in the external regions (see Sect.~\ref{sect:profile}), eight had already been studied by \citet{geller2015}, and the last star, S\,1053 is a double-lined spectroscopy binary according to \citet[][see Table~\ref{tab:bss}]{geller2015}. Seven of the objects in common with \citet{geller2015} were classified as ``single members'', two were marked as ``binary likely member'' one is the double-lined spectroscopy binary S\,1053. The remaining eight BSS candidates has a high membership probability for \textit{Clusterix} but not for \textit{UPMASK}. Five of them are labelled by \citet{geller2015}  as ``binary member'' or ``binary likely member''. In principle this can be explained by the fact that \textit{UPMASK} takes into account the uncertainties in the \textit{Gaia}~DR2 astrometric parameters which in principle may be larger for binaries. There is no additional information for the other three previously unstudied objects in the external regions of the cluster. Since there is no additional reason to prefer one over the other, we are going to analise the \textit{Clusterix} and \textit{UPMASK} BSS samples independently (BSSc and BSSu, respectively).

The stellar sample used in this work, including stellar identification, coordinates and membership probabilities derived by both \textit{UPMASK} and  \textit{Clusterix} is only available in electronic form
at the CDS.

%\begin{table}
%\centering
%\caption{M~67 features derived from GDR2.\label{table:m67_features}}
%\setlength{\tabcolsep}{0.7mm}
%\begin{tabular}{lccc}
%\hline
% & Mean & $\sigma$ & units \\
%\hline
%$\alpha_{GDR2}$ & 132.842$\pm$0.002 & 0.106$\pm$0.002 & deg\\
%$\delta_{GDR2}$ & 11.819$\pm$0.002 & 0.090$\pm$0.002 & deg\\
%$\mu_{\alpha*}$ & -10.972$\pm$0.005 & 0.211$\pm$0.006 & mas\,yr$^{-1}$ \\
%$\mu_{\delta}$ & -2.961$\pm$0.005 & 0.207$\pm$0.005 & mas\,yr$^{-1}$ \\
%$\varpi$ & 1.134$\pm$0.001 & 0.049$\pm$0.002 & mas\\
%\hline
%\end{tabular}
%\end{table}

\begin{table*}
\centering
\caption{Summary of BSSs}\label{tab:bss}
\setlength{\tabcolsep}{0.45mm}
\begin{tabular}{llcccccl}
\hline
ID$^a$ & ID$^b$ & Source\_id & distance & Class.$^c$ & Clusterix & UPMASK & Comments \\
\hline
1006 & S1066 & 604921202767814528 & 0.05 & (BL)M & M & M & SB1,RR,BSS\\
1007 & S1284 & 604918213470568576 & 0.06 & BM & NM & NM & SB1,RR,EX Cnc,PV,BSS\\
1010 & S977 & 604911478961895424 & 0.07 & (BL)M & M & NM & RR,BSS,SB1\\
1017 & S1466 & 604918831945837312 & 0.14 & (S)N & NM & NM & RR,CX109\\   
1020 & S751 & 604914189085426560 & 0.16 & SM & M & M & \\
1025 & S1195 & 604896566835438464 & 0.20 & BM & M & NM & SB1,RR,BSS\\
1026 & S1434 & 604903713660959488 & 0.21 & (BL)M & M & M & SB1,RR,BSS\\ 
2007 & S984 & 604911337227185920 & 0.05 & SM & M & NM & \\
2008 & S1072 & 604921374566321920 & 0.06 & BM & NM & NM & SB1,X37,CX24,YG\\    
2009 & S1082 & & 0.08 & BM& & & SB2,triple,ES Cnc,X4,CX3,PV,BSS\\ 
2011 & S968 & 604910310730777472 & 0.08 & SM & M & M & PV?,BSS\\
2013 & S1267 & 604918591427681024 & 0.10 & BM & M & NM &SB1,RR,BSS\\
2015 & S792 & 604921958681906176 & 0.12 & SM & M & M & \\    
2068 & S277 & 604984665204470016 & 0.56 & BM & NM & NM & SB1\\
3005 & S1263 & 604917835513456512 & 0.04 & SM & M & M & PV,BSS\\
3009 & S1273 & 604917934296937344 & 0.07 & SM & M & M & \\
3010 & S975 & 604911268507711232 & 0.07 & BM & NM & NM & SB1,RR,BSS\\      
3013 & S752 & 604911135364519808 & 0.10 & BM & NM & NM & SB1,RR,PV?,BSS\\   
4003 & S1036 & 604918041671889792 & 0.02 & (BL)M & M & NM & SB1,RR,W UMa,EV Cnc,X45,CX19,PV\\ 
4006 & S1280 & 604918179110923520 & 0.05 & (BL)M & M & M & SB1,RR,EW Cnc,PV,BSS\\
5005 & S997 & 604917285757663872 & 0.03 & BM & NM & NM & SB1,CX95,BSS\\	    
5071 & S145 & 598958684353074048 & 0.59 & SN & NM & NM & \\	    
6038 & S2226 & 604997206508853632 & 0.31 & SM & M & M & BSS \\
8006 & S2204 & 604917285757665920 & 0.04 & SM & M & M \\
9005 & S1005 & 604917560635575808 & 0.04 & BM & NM & NM & SB1\\	    
& & 602049342819274240 & 1.19 & & M & M & \\
3006 & S1053 & 604920923594172928 & 0.05 & BM & M & M & SB2\\
& & 578071536838119168 & 8.85 & & M & M & \\
& & 585923389890240128 & 9.37 & & M & NM & \\
& & 601873116017045376 & 2.24 & & M & NM & \\
& & 582587853864423040 & 7.04 & & M & NM & \\
\hline
\end{tabular}
\tablefoot{
\tablefoottext{a}{WIYN Open Cluster Study (WOCS) identifier.}
\tablefoottext{b}{Including red clump.}
\tablefoottext{c}{Membership classification from \citet{geller2015} (SM single member; SN single non-member; BM, binary member; BN binary non-member; BLM binary likely member).}
}
\end{table*}
    
\section{Radial density profile}\label{sect:profile}

The radial density profile is a basic tool to investigate the spatial distribution of the cluster stellar populations and their extension. To do that we have calculated the mean stellar surface density in concentric rings as $\rho_i={N_i}/{\pi(R^2_{i+1}-R^2_{i})}$  where $N_i$ is the number of stars in the $i-th$ ring with inner and outer radius $R_i$ and $R_{i+1}$, respectively. For this purpose we have used only the stars along the cluster sequence, excluding BSS candidates and binaries. The density uncertainty in each ring was estimated assuming Poisson statistics.

\begin{figure}
\centering
\includegraphics[scale=0.35]{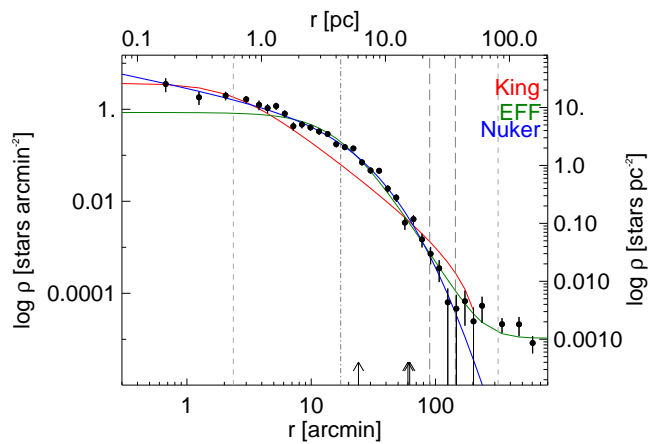}
\caption{Stellar radial density profile for M67 members (black filled circles). The different templates fitted have been plotted in different colours (see text for details). Vertical grey lines mark the derived $r_{\rm c}$ and $r_{\rm t}$ for \textit{King} (short dashed lines) and \textit{Nuker} (long dashed lines) templates and $r_{\rm c}$ for EFF template (dot-dashed line). Small arrows in the bottom show previous cluster size determinations available in the literature by \citet{bonatto2003}, \citet{davenport2010}, and \citet{gao2018} from left to right.}
\label{fig:densityprofile}
\end{figure}

The  radial density profile obtained is shown in Fig.~\ref{fig:densityprofile}. The stellar density decrease slowly with radius in the inner $\approx$30\arcmin. After this there is a steeper decrease till $\approx$150\arcmin. From there of the slope slows down but it does not get flattened since there are still members at larger distances which are not shown in the figure for clarity. The first conclusion is that there are member stars at distances larger than 100\arcmin~(25\,pc) from its centre. In fact a preliminary analysis of the \textit{Gaia}~DR2 sample showed that the cluster extends at least beyond $\sim$60\arcmin \, ($\approx$15\,pc) from its centre \citep{GaiaDR2_HRD}. The value found here is much larger than previous determinations in the literature. From 2MASS photometry, \citet{bonatto2003} found that the cluster extends up to a distance of about 24\arcmin\, ($\approx6$\,pc) while \citet{davenport2010}, using SDSS photometry, reported that the cluster extends to $\sim$60\arcmin \, ($\approx$15\,pc). More recently, \citet{gao2018} found a limiting radius of $\sim$62\arcmin ($\approx$16\,pc) using also \textit{Gaia}~DR2 but using a different method to assign membership probabilities.

%%%%%%%%FIGURE%%%%%%%%
%%%%%%%%FIGURE%%%%%%%%

In order to characterize the resulting density profile we have fitted it with three different analytic templates: \textit{King} \citep{King1962}; \textit{EFF} \citep{elson1987}; and \textit{Nuker} \citep{vandermarel2010}. Despite these expressions having been obtained empirically and not being the physical solutions of the collisionless Boltzamnn equation, they are useful to analyse the radial density profile. It is assumed that the density profile can be separated in three parts: core, bulk, and halo (or tidal debris). The different templates have been defined to focus on each of these parts. For example, \citet{King1962} tried to reproduce the bulk of the cluster, whereas \citet{elson1987} focused on the external tidal regions.

In the following we present the analytical templates of the density profiles investigated in this work. A more detailed explanation of each of them is provided by \citet{kupper2010MNRAS.401..105K}.

\begin{enumerate}
    \item The widely used {\sl King} template was presented by \citet{King1962}. It is in the form:
\begin{equation}
\rho(r)=\rho_{bg}+\rho_{0}\left[\dfrac{1}{\sqrt{1+(r/r_{\rm c})^2}}-\dfrac{1}{\sqrt{1+(r_{\rm t}/r_{\rm c})^2}}\right]^2
\end{equation}

\noindent where $\rho_{bg}$ is the background density and $\rho_{0}$ is the central density. The core radius $r_{\rm c}$ is defined as the distance between the centre and the point where $\rho(r)={\rho_{0}}/{2}$. The tidal radius, $r_{\rm t}$, is the position where $\rho(r)=\rho_{bg}$, the background density. The King template therefore consists of a flat core
and a bulk, but has no term for the tidal debris.
    \item Given that the outer parts of the density profiles of several clusters were not properly reproduced by King profiles,  \citet{elson1987}, {\sl EFF}, proposed this template which falls like a power-law without edge radius: 

\begin{equation}
\rho(r)=\rho_{0}\left[
1+\left(
\dfrac{r}{r_{\rm c}}\right)^2
\right]^{-\eta/2}
\end{equation}
\noindent where $\rho_{0}$, and $r_{\rm c}$ have the same meaning as in the case of the \textit{King} template. $\eta$ is the slope of the template for radii much larger than $r_{\rm c}$
%ground 
and $\eta$ is the slope of the template for radii much larger than the core radius.

\item The {\sl Nuker} template was initially proposed by \citet{lauer1995} to describe the surface brightness profiles of elliptical galaxies that showed a power-law cusp towards the centre and a logarithmic decline for radii larger than $r_{\rm c}$. An additional logarithmic slope from $r_{\rm t}$ was added by \citet{vandermarel2010} in order to describer the surface brightness of $\omega$~Centauri. This template is able to fit a cluster with a non-flat core, e.g. a core-collapsed cluster or a cluster with a massive central black hole. It has the form:

\begin{equation}
\rho(r)=\rho_{bg}+\rho_{0}2^{\frac{\eta-\gamma}{\alpha}}\left[\dfrac{r}{r_{\rm c}}\right]^{-\gamma}\left[1+\left(\dfrac{r}{r_{\rm c}}\right)^{\alpha}\right]^{-\frac{\eta-\gamma}{\alpha}}\left[1+\left(\dfrac{r}{r_{\rm r}}\right)^{\delta}\right]^{-\frac{\epsilon-\eta}{\delta}}
\end{equation}

\noindent where $\rho_{bg}$, $\rho_{0}$, and $r_{\rm c}$ are as before. $\gamma$ and $\eta$ are the power-law slopes inside and outside $r_{\rm c}$, respectively. The factor $\alpha$ smooths the transition between the two slopes. $r_{\rm t}$ is an additional break radii, from there off the density falls with $\epsilon$ logarithmic slope where $\delta$ determines the smoothness of the transition between $\epsilon$ and $\eta$.
\citet{vandermarel2010} found that the two break radii roughly correspond to the core an tidal radius of the cluster. In brief, this template consists of a flexible core, a bulk and a tidal debris.

\end{enumerate}

Each analytical template has been fitted to the observed profile using a non-linear fitting with the \textit{MPFIT} package \citep{mpfit}. The best fits in each case have been plotted with different colours in Fig.~\ref{fig:densityprofile}. The \textit{King} template %\citep[red line on Fig.~\ref{fig:densityprofile}, ][]{King1962},
(red line on Fig.~\ref{fig:densityprofile}),
consisting of a flat core an a bulk, is only able to reproduce innermost region, $\lesssim$3\arcmin. Up to $\approx$60\arcmin \, this template predicts less stars than observed.
The \textit{EFF} template \citep{elson1987}, shown as a green line in Fig.~\ref{fig:densityprofile}, is unable to reproduce the central region, up to $\approx$10\arcmin, predicting fewer stars than actually observed. On the contrary, for larger radii this template reproduces reasonably well the number of stars observed. The modified \textit{Nuker} template proposed by \citet{vandermarel2010} is in between of the previous ones. It is able to fit the data with a non-flat core, such as a core-collapsed cluster or a cluster with a massive central black hole. In the case of NGC~2682, this template (blue line in Fig.~\ref{fig:densityprofile}) reproduces the observed stellar density profile almost up to a radius of $\approx$200\arcmin. In the outermost radii the \textit{Nuker} template predicts a continuous logarithmic fall. However, the observed profile shows a change of the slope suggesting the existence of a tidal tail or a halo.

In order to check the statistical significance of the agreement, or disagreement, of the templates and the observed density profile we have computed the $\chi^2$ of the fit in the way
$$\chi^2=\sum\frac{(\rho_{i,t}-\rho_i)^2}{\sigma_{\rho_i}}$$
where $\rho_{i,t}$ is the density predicted by template $t$ at the $i-th$ ring. We obtained $\chi^2$= 6825.6, 71.5, and 2.4 for \textit{King}, \textit{EFF}, and \textit{Nuker} templates, respectively. The values for the best \textit{King} template fit are: $\rho_0=$3.8$\pm$0.6\,stars\,arcmin$^{-1}$; $r_{\rm c}$=2\farcm4$\pm$0\farcm2; and $r_{\rm t}$=319\arcmin$\pm$9\arcmin. In the case of \textit{EFF} template the best fit is obtained for: $\rho_0=$0.85$\pm$0.01\,stars\,arcmin$^{-1}$; $r_{\rm c}$=17\farcm27$\pm$0\farcm03; and $\eta$=4.23$\pm$0.01. Finally, the values of the best \textit{Nuker} template fit are: $\rho_0=$6.6$\pm$0.8$\times$10$^{-4}$\,stars\,arcmin$^{-1}$; $r_{\rm c}$=90\arcmin$\pm$2\arcmin; $\eta$=10.2$\pm$0.1; $\gamma$=0.57$\pm$0.01; $\alpha$=1.14$\pm$0.01; $r_{\rm t}$=144\arcmin$\pm$1\arcmin; $\delta$=28$\pm$4; and $\epsilon$=3.1$\pm$0.1. The \textit{Nuker} template reproduced better than the other the observed stellar density profile for NGC~2682 as the $\chi^2$ value confirms. 

%Finally, the \citet{kupper2010MNRAS.401..105K} template, KKBH, is a modification of the \citet{King1962} one allowing to reproduce the core with a power-law cusp. Therefore, it consists of a flexible core and a bulk but without halo/tidal debris term. This template (magenta line on Fig.~\ref{fig:densityprofile}) is able to reproduce most of the density profile. It predicts a higher density in the central region than other templates, but we do not have enough resolution to check its significance. Moreover, in the outermost regions, $\gtrsim$90-100\arcmin, this template predicts a slightly lower number of stars than observed. \RC{Give numbers for the best fit. Try KKBH with an additional term to account for the outermost region?} \textcolor{blue}{CJ: The outermost region in KKBH looks at least a factor 2 of understimation, isn't it ?}

\begin{figure*}
\centering
\includegraphics[scale=0.75]{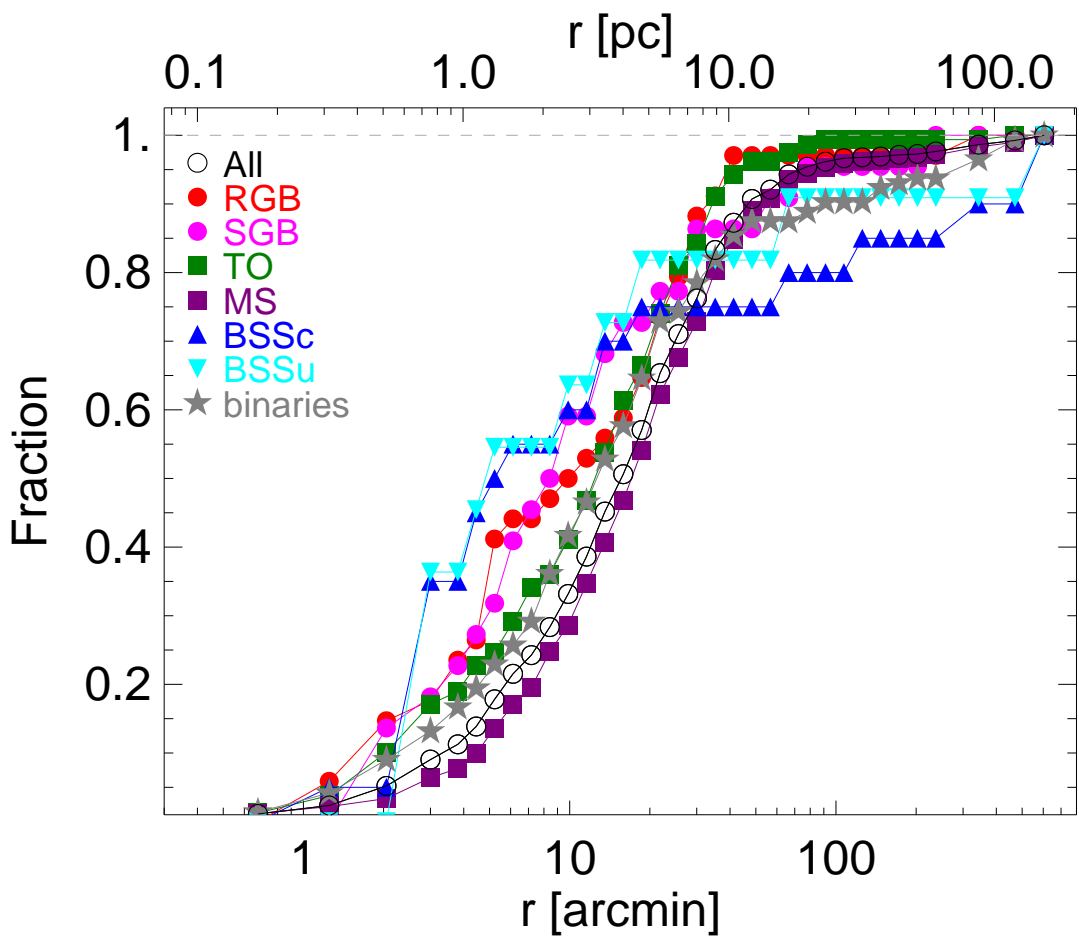}
\caption{Cumulative projected radial distribution of different populations identified in NGC~2682 normalised to the total number of stars in  each population. Errorbars have not been plotted for clarity.}
\label{fig:cumulative}
\end{figure*}

Figure~\ref{fig:cumulative} shows the cumulative projected radial distribution of the different NGC~2682 stellar populations normalised to the total number of the sub-population and including BSS and binaries. It is clear that the global distribution (open black circles) is mainly due to the MS stars (magenta squares) particularly in the outskirts. Turn-off objects (green squares) follow a similar trend than MS ones but they are more concentrated. While the 15\% of the MS population is located inside a radius, r$_{\rm 15\%}$, of 5\farcm6 the TO is concentrated inside 2\farcm7 (see Table~\ref{tab:population}). In fact, the r$_{\rm 15\%}$ increases when we go down in the cluster sequence. The same trend is observed for r$_{\rm 85\%}$, the radius that contains the 85\% of the stars of a given population. The binaries (grey stars in Fig.~\ref{fig:cumulative}) are slightly more concentrated that the MS objects as expected since they may be more massive objects. Since the mass of the star decreases as we go fainter along the cluster sequence the observed trend confirms that this cluster is mass-segregated with the most massive objects concentrated in the central regions. This behaviour has been reported in the literature by several authors \citep[e.g.][]{balaguer_nunez2007,geller2015,gao2018}.

The distribution of the BSS deserves a more detailed analysis. In principle, both \textit{Clusterix}, blue upward triangles in Fig.~\ref{fig:cumulative}, and \textit{UPMASK}, cyan downward triangles in Fig.~\ref{fig:cumulative}, samples shows a similar trend. The bulk of the BSS are confined in the central regions of the clusters with the 50\% of the population inside r$_{50\%}\sim$5\arcmin~(see Table~\ref{tab:population}). There is a fraction of BSSs at large radii although they need to be confirmed as really BSS objects. Between $\approx$10\arcmin~and 100\arcmin~there are almost no BSSs. This trend with the majority of the objects in the central region and several in the outskirts with almost none of them in between has been reported in globular clusters \citep[e.g.][]{ferraro1997,lanzoni2007}. The observed distribution could be due to the BSS formation mechanism as suggested by dynamical simulations \citep{mapelli2004}. The objects in the external regions would be the result of mass transfer in primordial binaries where BSSs in the cluster core have most likely a collisional origin. 

%%%%%%FIGURE%%%%%%%%
%%%%%%%%FIGURE%%%%%%%%

\begin{table}
\centering
\caption{Summary of cumulative projected radial distribution of different populations.\label{tab:population}}
\setlength{\tabcolsep}{0.7mm}
\begin{tabular}{lcccc}
\hline
Population & Nr Stars & r$_{15\%}$ & r$_{\rm 50\%}$& r$_{\rm 85\%}$\\
\hline
All$^{\rm a}$ & 808 & 4\farcm7 & 15\farcm6 & 37\farcm8 \\
RGB$^{\rm b}$ & 34 & 2\farcm1 & 9\farcm9 & 28\farcm4\\
SGB & 22 & 2\farcm3 & 8\farcm4 & 29\farcm4\\
TO & 158 & 2\farcm7 & 12\farcm5 & 30\farcm6\\
MS & 594 & 5\farcm6 & 17\farcm1 & 41\farcm5\\
BSSc & 20 & 2\farcm4 & 5\farcm2 & 237\farcm1\\
BSSu & 11 & 2\farcm4 & 4\farcm8 & 60\farcm1\\
Binaries & 144 & 3\farcm4 & 12\farcm7 & 40\farcm5\\
\hline
\end{tabular}
\tablefoot{
\tablefoottext{a}{Does not include BSS and binaries.}
\tablefoottext{b}{Including Red clump.}
}
\end{table}

%%%%%%%%FIGURE%%%%%%%%
\begin{figure*}
\centering
\includegraphics[scale=0.75]{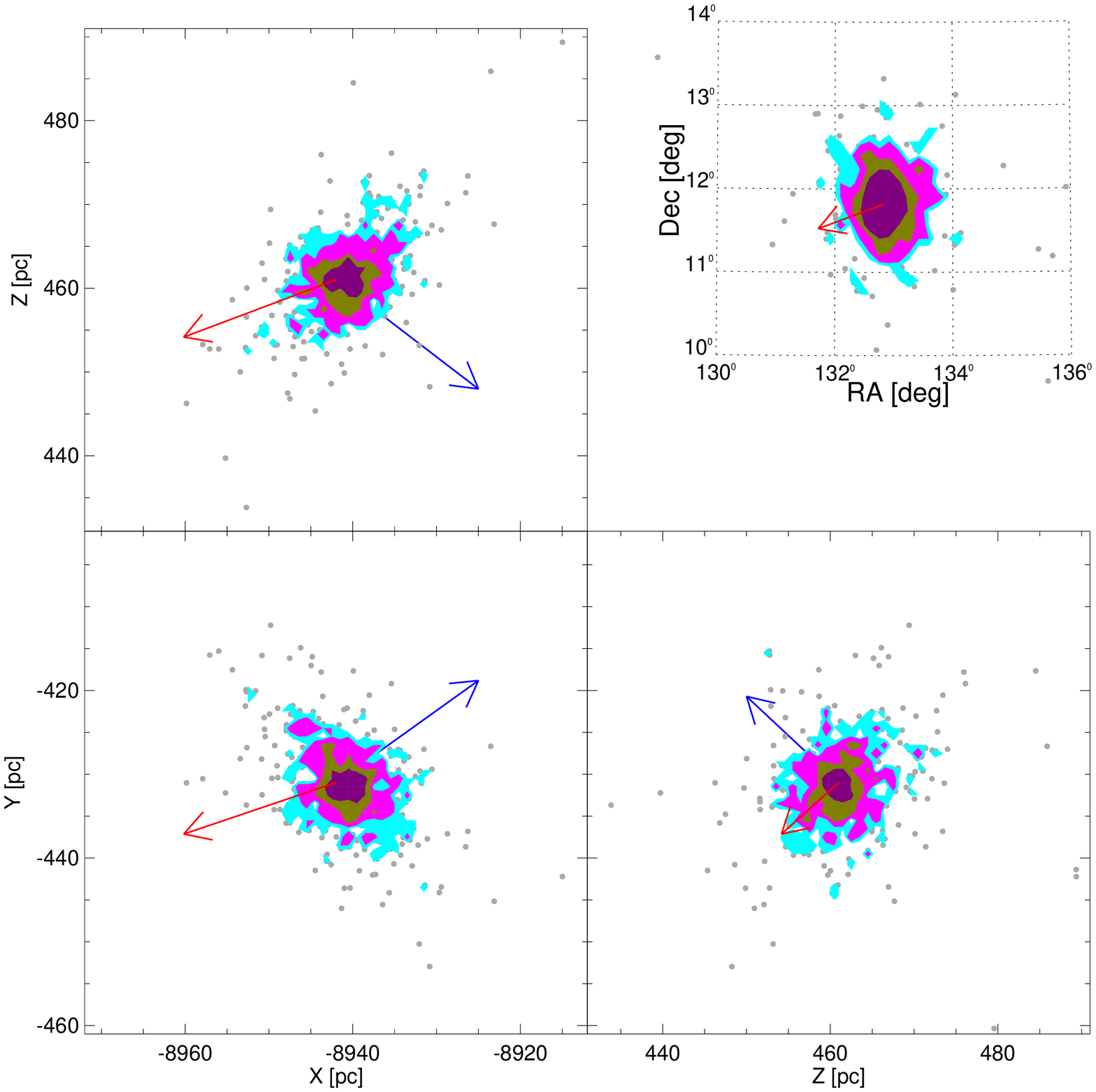}
\caption{Position of the M67 members in the Galactic Cartesian coordinates (grey dots). Contours show different stellar densities. Red arrows are proportional to the velocities in each axis derived by \citet{soubiran2018}. Blue arrows show the direction to the observer.}
\label{fig_xyz}
\end{figure*}
%%%%%%%%FIGURE%%%%%%%%

\section{Three-dimensional spatial distribution}
\label{ReverendBayes}
In order to further investigate the outskirts of NGC~2682 we have derived its 3D spatial distribution. Without observational uncertainties on the parallaxes the individual distances could be simply obtained as the inverse of the parallax, $d=\varpi^{-1}$. However, observational errors on measured parallaxes are not negligible and their direct propagation to distances through inversion may yield unacceptable results, that would translate into a strong elongation of the three-dimensional shape along the line of sight \citep[][]{1996MNRAS.281..211S, 2015PASP..127..994B, 2018A&A...616A...9L}. To overcome this difficulty, we used a Bayesian approach to parallax inversion described in Appendix~\ref{Bayes}, essentially following \cite{2015PASP..127..994B} but with a different prior, which combines two terms: the cluster proper, or core, and a field/halo term which accounts for stars that either form a diffuse envelope around the cluster or are serendipitously placed at a similar distance and share proper motions with the cluster. The first term is represented by a normal distribution for simplicity and the second term is the exponentially decreasing volume density prior introduced by \citet{2015PASP..127..994B}. A detailed analysis of the prior that is best for clusters is out of the scope of this paper. In our case the form of the prior does not have a big impact since the main result comes without 3D information, i.e. the sky-projected positions of stars which do not depends on parallaxes but shows a small tail in the opposite direction to that of the movement (top right panel of Fig.~\ref{fig_xyz}). On the other hand, if the parallaxes are giving us enough information then the shape of prior does my not make a large difference unless it is very unrealistic. %} \footnote{\MP{The text was not fully correct. My prior has two terms: one is a Gaussian centered on the clusters distance and the other is the conventional exponentially decreasing volume density prior introduced by Bailer Jones 2015 Sect. 7 \href{https://arxiv.org/pdf/1507.02105.pdf}{astro-ph:1507.02105} The Gaussian part is meant to represent the cluster, the field part the field. I understand that there should be no field thanks to GAIA's pm cleaning but it makes sense that still some stars are misclassified. It turned out that without the field prior (using Gaussian only) the distribution would look unrealistic. While discussing the specific form of the prior that is best for clusters is very interesting and based on my limited knowledge of the literature seems an open problem, my understanding was that for this paper the form of the prior is not very important, as the main result comes without 3D information, i.e. by just looking at the sky-projected positions of stars and ignoring parallaxes. If you want to write a paper specifically on the prior for clusters I would be totally happy to to that. On the other hand if the parallaxes are giving us enough information then the prior should not matter much; if the end result depends too much on the prior then it means that the data we have are not good enough. This is definitely not the case with GAIA data, so the prior should not make too much of a difference, unless it is very unrealistic}}. 

The distribution of NGC~2682 in Galactic cartesian coordinates is shown in Fig.~\ref{fig_xyz}. In this coordinate system, the $X$ axis points towards the Galactic centre, $Y$ is positive towards the local direction of rotation in the plane of the Galaxy, and $Z$ points towards the North Galactic Pole. The Galactic centre is the origin of this system with the Sun located at (-8340,0,27)\,pc \citep[see][and references therein]{katz18}. The spatial distribution of the stars is not isotropic in the $XZ$- and $ZY$-planes despite the symmetric nature of the normal term representing the cluster in the prior. This is a clear indication that the observed parallaxes have a non-negligible effect on the posterior.
In the case of the $XZ$-plane (top left panel), there is a clear tail almost in the direction opposite to that of the cluster's movement. An elongation is observed in the direction of the movement in the $ZY$-plane (bottom right panel). A small elongation is also observed in the $XY$-plane (bottom left panel), but in this case in the direction perpendicular to the movement.

\citet{davenport2010} using ground based photometry reported an asymmetric halo up to a distance of $60$\arcmin~ from the cluster centre elongated roughly in-line with the proper-motion vector on the sky plane \citep[see Fig~10 of][]{davenport2010}. However, such elongation is not observed in our case (top right panel of Fig.~\ref{fig_xyz}). However, as commented before a small tail is visible in the opposite direction to that of the movement.

\section{Physical interpretation}
\label{physics}
We estimate the radius of NGC~2682's Hill sphere $R_{\rm H}$ in the Galactic potential by approximating the latter to a point-particle potential, obtaining 
\begin{equation}
\label{Hillr}
    R_{\rm H} = D \times {\left(\frac{m}{3M} \right)}^{1/3}
\end{equation}
where $D$ is the distance of NGC~2682 from the Galactic centre, $m$ the mass of NGC~2682, and $M$ the mass of the Galaxy enclosed within NGC~2682 orbit.
The quantity we know with less uncertainty is $D$, which is estimated as $D$ = 8.9\,kpc from \textit{Gaia}~DR2 \citep{tristan2018b}
On the other hand, the total mass of the Galaxy is model dependent, e.g. $M = 10^{12}$ $M_\odot$ \citep[][]{2017MNRAS.465...76M} and this uncertainty is compounded by the fact that we want to consider only the mass within NGC~2682 orbit. Similarly, there is likely an uncertainty of a factor $2$ in estimates of NGC~2682 mass such as \cite{hurley2005} which sets it at $m = 2 \times 10^3$ $M_\odot$. Luckily, the dependence to the power of $1/3$ in Eq.~\ref{Hillr} ensures that the uncertainty on these numbers has little effect on $R_{\rm H}$. 
%\textcolor{blue}{CJ: How does this number been calculated ? from the sample in this paper ? The sample of members is limited to G<18, so the mass can be underestimated. In the Gaia Collaboration paper there are 1500 members down to G=20 being 1000 of them brighter than 18 and 500 of them fainter than 18mag}\MP{see answer below}
%\textcolor{green}{MM: $\sim{}10^{12}$ M$_\odot$ is the total mass of the Milky Way, but I guess  we have assumed that this is the mass inside $D = 10.7$ kpc. This might lead to a significant overestimate, which might affect our main result by a few. Sorry for not reading this before. Mario, have you checked this?}\MP{see answer below}
To illustrate this, let us consider two different extreme scenarios:
\begin{itemize}
\item Overestimate the enclosed Galactic mass at $M = 10^{12}$ $M_\odot$ and underestimate NGC~2682 mass at $m = 10^3$ $M_\odot$, obtaining $R_{\rm H} = 6.2$ pc
\item Underestimate the enclosed Galactic mass at $M = 10^{11}$ $M_\odot$ and overestimate NGC~2682 mass at $m = 10^4$ $M_\odot$, obtaining $R_{\rm H} = 28.6$ pc
\end{itemize}
In both cases, our observations show that M67 stars can be found further out than $R_{\rm H}$, suggesting that they are indeed extra-tidal. Such loosely bound stars may have originated as the consequence of internal relaxation (i.e. they may be dynamically evaporated stars) or as the effect of an external disturbance.

%With these numbers we obtain $R_{\rm H} = \mu^{1/3} 9.3$ pc.
%\textcolor{red}{AV: you should give some references how you calculate $\mu$. This can affect the results}
%\MP{Let me answer here to both Ricardo and Michela: I just patched the issue of having an underestimated NGC~2682 mass (pointed out by Ricardo) and overestimated Milky Way mass (pointed out by Michela) by adding a fudge factor and showing that it does not really affect the results because it is to the power of $1/3$; this may be enough if we are in a rush to submit, but if you think it looks sloppy then I can calculate the mass of the MW properly using a model... at that point I think that it would be worth to use the actual potential instead of a point particle potential though.}

%While this calculation is likely to break down near the Galactic disk, where the potential is quite different from that for a point mass,

\begin{figure}
\centering
\includegraphics[scale=0.38]{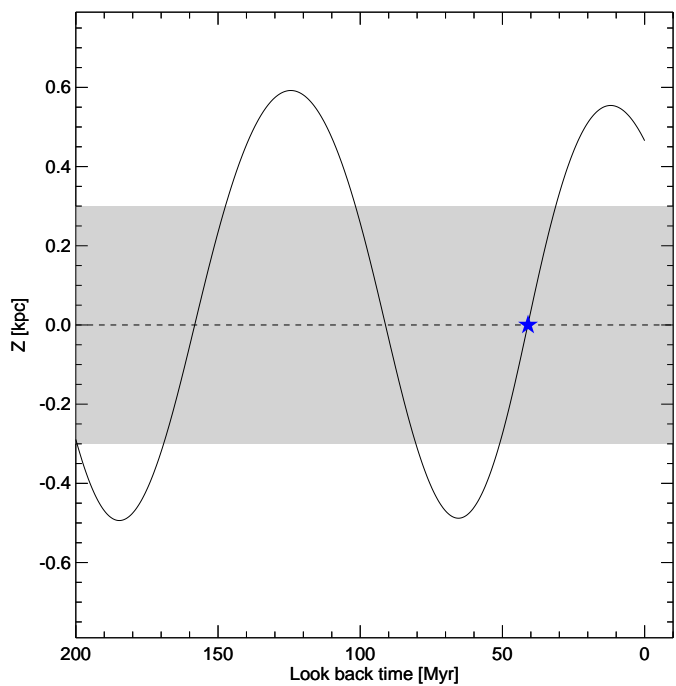}
\caption{Height on the Galactic plane $Z$ as a function of time in the last $200$\,Myr of the orbit of NGC~2682 as integrated by \textit{GRAVPOT16}. The green star shows the time of the last disk passage $\approx$40\,Myr ago. The shadow area shows the extension of the thin disk.}
\label{LastDiskPassage}
\end{figure}

We used the software \textit{GRAVPOT16} \citep[][]{GravPot16Thesis,2012A&A...538A.106R, 2003A&A...409..523R} to integrate the orbit of NGC~2682 back in time for 1\,Gyr. We find that the last disk passage of NGC~2682 happened $\approx$40\,Myr ago, as shown in Fig.~\ref{LastDiskPassage} which plots the height over the Galactic plane $Z$ as a function of time in the last 200\,Myr. The shadow region shows the width of the thin disk according to \citet{juric2008}. Note that the thick disk width is larger than the area plotted. This result is qualitatively compatible to the findings of \citet{1538-3881-143-3-73}, even though the exact time of last crossing differs due to systematics introduced by the usage of different Galactic potential models and to the lack of good proper motions in the pre-{\it Gaia} era.

According to \cite{hurley2005} the relaxation time of a realistic $N$-body model of NGC~2682 was initially (before mass loss) $\approx$300\,Myr . Even allowing for mass loss shortening the relaxation time by a factor ten, it is reasonable to assume that the effects of this disk passage have not yet been erased by dynamical relaxation, especially in the external regions of the clusters where the relaxation time is shorter than the half-light relaxation time. According to Fig.~\ref{LastDiskPassage} NGC~2682 has had three crossing disk passages in the last 200\,Myr which implies that the cluster is constantly disturbed and it will never be relaxed because the frequent passages.

The extra-tidal features we observe in M67 would thus be quite a natural result of disk shocking.
This is to be confirmed quantitatively with dedicated $N$-body models in an upcoming paper (Pasquato et al. 2019, in preparation).

An estimate of the effects of disk crossing can be obtained in the analytical approximation of an impulsive shock \citep[][]{1972ApJ...176L..51O}, where every star gets a nudge in velocity
\begin{equation}
    \delta v_z = \frac{4 \pi G \Sigma z}{V}
\end{equation}
where $V$ is the velocity with which the cluster crosses the disk, $z$ is the distance along the $Z$ coordinate of a given star from the cluster centre, and $\Sigma$ is the density per unit surface of the Galactic disk.
This corresponds, for a typical star moving with velocity $v = GM/R_{\rm h}$ relative to the cluster centre, to a relative change in energy of order
\begin{equation}
    \frac{\delta E}{E} = \frac{{\delta v_z}^2}{v^2} \approx \frac{16 \pi^2 G \Sigma^2 R_{rm h}^3}{M V^2}
\end{equation}
where $R_{\rm h}$ is the cluster's half-mass radius, and $M$ the total mass of the cluster. This is proportional to $M^2_D\sigma^2/M^2V^2$ where $M_D$ is the mass enclosed, on the galactic disk, by a circumference with radius the cluster's half-mass radius, and $\sigma$ is the cluster's velocity dispersion.

Based on our orbital calculations, $V \approx 30$ km s$^{-1}$, and a surface mass density of $30$ $M_\odot$ pc $^{-2}$ \citep[][]{2016ApJ...816...42M} we obtain $\delta E/ E \approx 0.3$, which corresponds to an expansion of the virial radius of the cluster of the same quantity, i.e.  $\delta R/ R \approx \delta E/ E \approx 0.3$. This is enough to justify the presence of a relatively large number of extratidal stars. %\MP{DONE devo controllare i conti}

%%%%%%%%FIGURE%%%%%%%%
%%%%%%%%FIGURE%%%%%%%%

\section{Conclusions}

Two different methods, \textit{Clusterix} and \textit{UPMASK} have been used to assign membership probabilities to stars in the line of sight of NGC~2682 up to a distance of 150\,pc (10 degrees) using the information provided by the \textit{Gaia}-DR2 catalogue. Additionally, we have estimated distances to obtain three-dimensional stellar positions using a Bayesian approach to parallax inversion, with an appropriate prior for star clusters. The main results of our analysis are:

\begin{itemize}
    \item The cluster extends up to 200\arcmin~(50\,pc) which implies that its size is at least twice as previously believed. This exceeds the cluster Hill sphere based on the Galactic potential at the distance of NGC~2682.
    \item Only a template with a power-law cusp towards the centre and a logarithmic decline for larger radii is able to reproduce the observed stellar density profile. In any case, there are stars outside the tidal radius.
    \item As reported in previous studies, we confirm that the cluster is mass-segregated with the most massive objects concentrated in the central regions.
    \item As observed in several globular clusters the BBSs show a bimodal distribution with the majority of the object located in the core an several objects in the outskirts without anything in between. Although the objects at larger radii need to be confirmed as real BSSs.
    \item The extra-tidal stars in NGC~2682 may originate from external perturbations such as disk shocking or dynamical evaporation from two-body relaxation. The former origin is plausible given the orbit of NGC~2682, which crossed the Galactic disk $\approx 40$\,Myr ago. We plan to further investigate this scenario with direct N-body models.
\end{itemize}

\begin{acknowledgements}

This work has made use of data from the European Space Agency (ESA) mission
{\it Gaia} (\url{https://www.cosmos.esa.int/gaia}), processed by the {\it Gaia}
Data Processing and Analysis Consortium (DPAC,
\url{https://www.cosmos.esa.int/web/gaia/dpac/consortium}). Funding for the DPAC
has been provided by national institutions, in particular the institutions
participating in the {\it Gaia} Multilateral Agreement.
This project has received funding from the European Union's Horizon $2020$
research and innovation programme under the Marie Sk\l{}odowska-Curie grant agreement No. $664931$
This work was partly supported by the MINECO (Spanish Ministry of Economy) through grant ESP2016-80079-C2-1-R (MINECO/FEDER, UE) and MDM-2014-0369 of ICCUB (Unidad de Excelencia 'María de Maeztu'),  European Community's Seventh Framework Programme (FP7/2007-2013) under grant agreement GENIUS FP7 - 606740 and the European Commission Framework Programme Horizon 2020 Research and Innovation action under grant agreement ASTERICS 653477. Support from PREMIALE 2016 MITiC is acknowledged. Based on Clusterix 2.0 service at CAB (INTA-CSIC). DB is supported in the form of work contract FCT/MCTES through national funds and by FEDER through COMPETE2020 in connection to these grants: $UID/FIS/04434/2019$; $PTDC/FIS-AST/30389/2017 \& POCI-01-0145-FEDER-030389$

\end{acknowledgements}

\bibliographystyle{aa} 
\bibliography{biblio_m67_spatial.bib}

\appendix 

\section{Clusterix membership selection}\label{apen:clusterix}

\textit{Clusterix} is a web-based tool for the determination of membership probabilities from proper motion data using a non-parametric approach based on the formalism described by \citet{galadi1998}, in a new implementation whose details can be found in \citet{clusterix}. In the sky area occupied by the cluster, the frequency function is assumed to be made up from two contributions: cluster and field stars. The tool performs an empirical determination of the frequency functions from the vector-point diagram (VPD) without relying on any previous assumption about their profiles. To disentangle the two populations, {\it Clusterix} studies the VPD corresponding to two areas on the plane of the sky: one in the cluster core and another one far enough from the cluster to have a small (negligible) cluster contribution, but close enough to still be a good representation of the field in the cluster area. \textit{Clusterix} allows to search in an interactive and visual way the appropriate areas until an optimal separation is obtained (more details will be given in Balaguer-Núñez et al, in preparation). Several limits can be adjusted to make the calculation computationally feasible without interfering in the quality of the results: in this case we have set a constrain in proper motions (-15$\leq\mu_{\alpha*}\leq$+15 and -15$\leq\mu_{\delta}\leq$+15\,mas\,yr$^{-1}$), proper motions errors (below 5\,mas\,yr$^{-1}$) and magnitude ($G<$18). Given that the \textit{Clusterix} selection is only based on proper motions we added an additional constrain in parallaxes choosing only those objects within $\pm2\sigma$ from the median of the $\varpi$ distribution. The current version of \textit{Clusterix} does not apply any criteria based on spatial distribution nor photometry for membership assessment. 

There is no rigorous way to decide where to set the limit between members and non-members in a list sorted by membership probabilities. This non-parametric membership method produces an expected number of cluster members that serves as an indication of the probability limit for the most probable cluster members. However, when the study is done over a large sky area, as in this case, the field surrounding the cluster cannot be considered perfectly homogeneous, due to the existence of other structures in the background and surroundings (e.g. other clusters). This is not a problem for the calculation of the membership, as long as the area chosen to represent the cluster is not too far from the cluster core, but it has an impact on the calculation of the expected number of cluster members, as this quantity depends on the total number of stars in the area. In this case, we have performed this study in two steps: first considering an area of 5\,deg where the field is homogeneous enough and the number of expected cluster members gives a reliable result, followed by a second step, using that list of members (where most of the cluster is found) to fix the membership probability cut on the study of the 10\,deg area ($p\geq$0.81).

\section{Bayesian inference of stellar distances for reconstructing the 3D shape of NGC~2682}
\label{Bayes}

As commented in Sect.~\ref{ReverendBayes} the propagation of the non negligible $\varpi$ uncertainties causes that the determination of distances through the inversion of the parallax, $d=\varpi^{-1}$, produces erroneous results. In order to overcome this issue we have used a Bayesian approach to determine distances, $d$, to the objects analysed in this paper. To do that we follow \citet{2015PASP..127..994B} Bayesian approach, using for each star the parallax $\varpi$ to update a prior distribution in distance $\Pi(d)$, obtaining an (un-normalized) posterior distribution as follows:
\begin{equation}
\label{Bayeseq}
    f(d | \varpi) \propto f(\varpi | d) \Pi(d)
\end{equation}
where
\begin{equation}
    f(\varpi | d) \propto e^{-{(\varpi - \frac{1}{d})}^2/2 \sigma^2_{\varpi}}
\end{equation}
with $\sigma^2_{\varpi}$ equal to the quoted error bar on each parallax measurement, i.e. we assumed that uncertainties on the parallax are Gaussian, with a standard deviation equal to the quoted error bar $\sigma_{\varpi}$.
Regarding the prior, we used a Gaussian centred at the average distance of NGC~2682, $d_0\approx$860\,pc \citep{tristan2018b}
to represent the cluster, plus a term that accounts for field stars or more in general stars that may be forming a diffuse envelope around NGC~2682, for which we do not wish to make overly restrictive assumptions. The standard deviation of the Gaussian term, $\sigma_d$ was chosen equal to the standard deviation of physical distances of member stars projected on the plane of the sky. This is equivalent to assuming, in the prior, that the cluster term is spherical. While the choice of a Gaussian prior may seem unjustified from the physical point of view we chose it for simplicity, as it is significantly easier to implement than e.g. a King model. The resulting prior is as follows:
\begin{equation}
    \Pi(d) = \alpha \xi(d) + (1 - \alpha) \frac{1}{\sqrt{2 \sigma_d}} e^{-{(d-d_0)}^2/2\sigma^2_d}
\end{equation}
where $\xi(r)$ is a (normalized) exponentially decreasing volume density prior \citep[see Eq.~17 of ][with $L$=8\,kpc]{2015PASP..127..994B}, and $1 - \alpha$ is the {\it UPMASK} membership score. As the \cite{2015PASP..127..994B} prior is essentially a smoothly truncated constant density prior it is a good choice for not making strong assumptions about the start that happen to be loosely associated to NGC~2682.
The adopted distance for each star is estimated with the mode of the posterior distribution. As pointed out in \citet{2018A&A...616A...9L}, the whole posterior distribution should be kept, rather than summarized, as the end result of a correct Bayesian inference procedure. This applies in our case a fortiori, because we are not interested in the position of individual stars as much as we are interested in the overall mass density distribution of the whole cluster. However in this letter, for the sake of concreteness, we decided to obtain anyway a point estimate from the posterior distribution. %i.e.
%\begin{equation}
%\label{norma}
%r_{\texttt{adop.}} = \frac{\int_{-\infty}^{+\infty} r f(r | p) dr}{\int_{-\infty}^{+\infty} f(r | p) dr}.
%\end{equation}
%The integrals have been evaluated numerically for each star. The approach in Eq.~\ref{norma} is justified because any normalization constant in $f(r | p)$ simplifies away.
\end{document}